 \documentclass[11pt]{article}
\input{amssym.def}
\input{amssym}
\newtheorem{theorem}{Theorem}
\newtheorem{proposition}[theorem]{Proposition}
\newtheorem{lemma}[theorem]{Lemma}

\def\D{{\mathcal D}}
\def\H{{\mathcal H}}

\def\Q{{\mathcal Q}}
\def\PT{{\mathcal {PT}}}
\def\T{{\mathcal {T}}}
\def\R{{\mathcal R}}

\def\H{{\mathcal H}}  

\def\R{\Bbb R} 
\def\Z{\Bbb Z} 
\def\N{\Bbb N}
\def\T{{\mathcal  T}}  
\def\C{\Bbb C}

\def\Sc{Schr\"o\-din\-ger}

\def\la{\langle}
\def\be{\begin{equation}}
\def\ee{\end{equation}}
\def\bea{\begin{eqnarray}}
\def\eea{\end{eqnarray}}
\def\ra{\rangle}
\def\ds{\displaystyle}

\def\ep{\epsilon}
\def\RSPE{Ray\-leigh-\Sc\ per\-tur\-ba\-tion ex\-pan\-sion}

\begin{document}
\baselineskip=19pt
\begin{center}
{\large\bf\sc  A criterion for the reality of the spectrum of $\PT$ symmetric \Sc\ operators with complex-valued periodic potentials}
\end{center}
\vskip 13pt
\begin{center}
 Emanuela Caliceti and Sandro Graffi
 \\
 {\small Dipartimento di Matematica, 
Universit\`a di Bologna, 40127 Bologna, Italy 
\footnote{caliceti@dm.unibo.it, graffi@dm.unibo.it}}
\end{center}
\begin{abstract}
\noindent
Consider in $L^2(\R)$   the  \Sc\ operator family
$H(g):=-d^2_x+V_g(x)$ depending on the real parameter $g$, where $V_g(x)$ is a complex-valued but $PT$ symmetric periodic  potential.   An explicit condition on $V$ is obtained  which ensures that the spectrum of $H(g)$ is purely real and band shaped; furthermore, a further condition is obtained which ensures that the spectrum contains at least a pair of complex analytic arcs. 
 
\end{abstract}
\vskip 1cm   
%
 
%

\section{Introduction and statement of the results}
\setcounter{equation}{0}%
\setcounter{theorem}{0}%
There is currently an intense and ever increasing activity on an aspect of quantum theory known as 
$PT$-symmetric quantum mechanics (see e.g.\cite{Be1}, \cite{Be2}, \cite{Be4}, \cite{BBJ}, \cite{Cn1}, \cite{Cn3}, \cite{Zn1}, \cite{Sp}, \cite{Cn2}). Mathematically speaking, in the simplest, one -dimensional case one deals with  the    stationary
\Sc\ 
 equation  
  \be
 \label{S}
 H\psi:=(-\frac{d^2}{dx^2}+V)\psi=E\psi\;.
 \ee
 where  the potential $V(x)$ can be  complex-valued  but is invariant under the 
combined action of 
the linear  parity operation $P$, $P\psi(x)=\psi(-x)$, and of the  (anti)linear   "time-reversal"  symmetry, i.e. the 
complex-conjugation operation $T\psi(x)=\overline{\psi}(x)$; namely, 
$\overline{V}(-x)=V(x)$. The basic mathematical problem is to determine under what conditions, if any,  on the 
complex $PT$ symmetric  potential $V$ the 
spectrum of the corresponding \Sc\ operator is purely real. 

Here we deal with this problem in the context of periodic potentials on $\R$, already considered in \cite{Ah}, \cite{BDM},\cite{Ce}, \cite{CR}, \cite{Jo}, \cite{Shin}. Without loss, the period
is assumed to be $2\pi$.  If $V$ is periodic and real valued it is well known (see e.g. \cite{BS}) that, 
under mild regularity assumptions, 
the spectrum is absolutely continuous on $\R$ and band shaped. 

It is then natural  to ask whether or not there exist  classes of $PT$ symmetric, complex  periodic 
potentials generating \Sc\ operators with real band spectrum. This question  has been examined in \cite{Ah}, \cite{BDM},\cite{Ce}, \cite{CR}, \cite{Jo}, by a combination of numerical and WKB techniques, in several particular examples.   It was later proved in \cite{Shin} that the above arguments  cannot exclude the occurrence of complex spectra, and actually 
 a condition has been isolated under which $H$ admits complex spectrum consisting of a disjoint  union of 
analytic arcs \cite{Shin}. \par
Our first result is the explicit determination a class of $PT$ symmetric, complex  periodic 
potentials admitting real band spectrum.  Denote $\T(u)$ the non-negative quadratic form in $L^2(\R)$  with domain $H^1(\R)$ defined by the kinetic energy:
\be
\label{T}
\T(u):=\int_\R |u^\prime|^2\,dx, \qquad u\in H^1(\R)
\ee
Let $q$ be  a real-valued, tempered distribution.  Assume: 
\begin{itemize}
\item[(1)] $q$ is a $2\pi$-periodic, $P$ symmetric distribution  belonging to $H^{-1}_{loc}(\R)$; 
\item[(2)] $W(x):\R\to \C$ be\-longs  to $L^\infty(\R)$ and is  $PT$-symmetric, $\overline{W(-x)}=-W(x)$, . 
\item [(3)]  $q$ generates a real quadratic  form $\Q(u)$  in $L^2(\R)$ with domain $H^1(\R)$;
\item [(4)] $\Q(u)$ is relatively bounded with respect to $\T(u)$ with relative bound $b<1$, i.e. there are $b<1$ and  $a>0$ such that
\be
\label{rb}
\Q(u)\leq b T(u)+a\|u\|^2
\ee
\end{itemize}
Under these assumptions the  real quadratic form
\be
\label{Q0}
\H_0(u):=\T(u)+\Q(u) \quad u\in H^{1}(\R)
\ee
is closed and bounded below in $L^2(\R)$. We denote $H(0)$ the corresponding self-adjoint operator. This is the self-adjoint realization of the formal differential expression (note the abuse of notation)
$$
H(0)=-\frac{d^2}{dx^2}+q(x)
$$
Under these circumstances it is known  (see e.g. \cite{AGHKH}) that the spectrum of ${H}(0)$  is continuous and  band shaped. For $n=1,2,\ldots$ we denote
 $$
 B_{2n}:=[\alpha_{2n},\beta_{2n}], \quad   B_{2n+1}:=[\beta_{2n+1},\alpha_{2n+1}]
 $$ 
 the bands of ${ H}(0)$, and $\Delta_n:=]\beta_{2n},\beta_{2n+1}[$, $]\alpha_{2n+1},\alpha_{2(n+2)}[$  the gaps between the bands. Here: 
$$
0\leq \alpha_0\leq \beta_0\leq \beta_1\leq \alpha_1\leq \alpha_2\leq \beta_2\leq\beta_3\leq\alpha_3\leq\alpha_4\leq\ldots. 
$$
The maximal multiplication operator by $W$ is continuous in $L^2$, and therefore so is the quadratic form $\la u,W u\ra$. It follows that the quadratic form family
\be
\label{Q}
\H_g(u):=\T(u)+\Q(u)+g \la u,Wu\ra, \quad u\in H^{1}(\R)
\ee
is closed and sectorial in $L^2(\R)$ for any $g\in\C$.  
\par\noindent
We denote  ${H}(g)$ the uniquely  associated $m-$ sectorial operator in $L^2(\R)$. This is the realization of the formal differential operator family
$$
H(g)=-\frac{d^2}{dx^2}+q(x)+gW(x)
$$
 By definition, ${ H}(g)$ is a holomorphic  family of operators of type B in the sense of Kato for $g\in\C$; by (1) it is also $PT$ symmetric for $g\in\R$. Our first result deals with its spectral properties. 
\begin{theorem}
\label{t1}
Let all gaps of $H(0)$ be open,   namely: $ \alpha_n < \beta_n < \alpha_{n+1}$ $\forall \,n\in\N$, and let there exist $d>0$ such that
\be
\label{infgap}
\frac12\inf_{n\in\N} \Delta_n:=d>0
\ee
Then, if 
\be
\label{rc}
|g|<\frac{d^2}{2(1+d)\|W\|_\infty}:=\overline{g}
\ee
there exist 
$$
0 \leq \alpha_0(g)< \beta_0(g)< \beta_1(g)<\alpha_1(g)<\alpha_2(g)<\beta_2(g)\ldots
$$
such that
\be
\sigma({H}(g))=\left( \bigcup_{n\in\N}B_{2n}(g)\right)\bigcup \left( \bigcup_{n\in\N}B_{2n+1}(g)\right)
\ee
where, as above
$$
B_{2n}(g):=[\alpha_{2n}(g),\beta_{2n}(g)], \quad B_{2n+1}(g):=[\beta_{2n+1}(g), \alpha_{2n+1}(g)].
$$
\end{theorem}
{\bf Remark} 
\newline
The theorem states that for $|g|$ small enough the spectrum of the non self-adjoint operator $H(g)$  remains real and band-shaped. The proof is critically dependent on the validity of the lower bound (\ref{infgap}). Therefore it cannot apply to smooth potentials $q(x)$, in which case the gaps vanish as $n\to\infty$. Actually we have the following
\par\noindent
{\bf Example}
\newline
A locally $H^{-1}(\R)$ distribution $q(x)$ fulfilling the above conditions is 
$$
q(x)=\sum_{n\in\Z}\delta(x-2\pi n)
$$
the periodic $\delta$ function.  Here we have:
$$
\Q(u)=\sum_{n\in\Z}|u(2\pi n)|^2=\int_\R q(x)|u(x)|^2\,dx,\quad u\in H^1(\R)
$$
This example is known as the Kronig-Penney model in the one-electron theory of solids.  Let us verify that condition (\ref{rb}) is  satisfied. As is known,  this follows from the  inequality (see e.g. \cite{Ka}, \S VI.4.10):
\begin{eqnarray*}
 |u(2\pi n)|^2\leq \epsilon\int_{2\pi n}^{2\pi (n+1)}|u^\prime (y)|^2\,dy+\delta \int_{2\pi n}^{2\pi (n+1)}|u (y)|^2
\end{eqnarray*}
where $\epsilon$ can be chosen arbitrarity small for $\delta$ large enough.  In fact, if $u\in H^1(\R)$ this inequality yields:
$$
 \int_{\R}q(x)|u(x)|^2\,dx=\sum_{n\in\Z}|u(2\pi n)|^2\leq \epsilon\int_\R |u^\prime (y)|^2\,dy+\delta \int_{R}|u (y)|^2
$$
$$
=\ep \T(u)+\delta \|u\|^2, \qquad u\in H^1(\R)
$$
which in turn entails the closedness of $\T(u)+\Q(u)$ defined on $H^1$ by the standard Kato criterion. The closedness and sectoriality of $\H_g(u)$ defined on $H^1(\R)$ is an immediate consequence of the continuity of $W$ as a maximal multiplication operator in $L^2$. For the verification of (\ref{infgap}), see e.g.\cite{AGHKH}.  Hence any bounded $PT$-symmetric periodic perturbation of  the Kronig-Penney potential has real spectrum for $g\in\R$, $|g|<\overline{g}$, where $\overline{g}$ is defined by (\ref{rc}). .
\vskip 0.3cm
As a second result, we  show that an elementary argument of perturbation theory 
allows us to sharpen the result of \cite{Shin} about the  existence of complex spectra for $PT$-symmetric periodic potentials. 
\par\noindent
Let indeed $W(x)\in L^\infty(\R;\C)$ be a $2\pi$-periodic function. Then the continuity of $W$ as a multiplication operator in $L^2(\R)$  entails that the  \Sc\ operator
\be
\label{K}
{ K}(g)u:= -\frac{d^2u}{dx^2}+gWu,\quad u\in D( K):=H^2(\R), \quad g\in\R
\ee
is closed and has non-empty resolvent set.  Consider the  Fourier expansion of $W$:
$$
W(x)=\sum_{n\in\Z}w_ne^{inx},\quad w_n=\frac{1}{2\pi}\int_{-\pi}^\pi\,W(x)e^{-inx}\,dx
$$
which converges pointwise almost everywhere in $[-\pi,\pi]$.  Then we have
\begin{theorem}
\label{complesso}
Let  $W(x)$ be $PT$-symmetric, namely $\overline{W(-x)} =W(x)$. Then
\begin{enumerate}
\item $\overline{w}_n=w_n$, $\forall\,n\in\Z$;
\item Furthermore, let there exist $k\in\N$, $k$ odd, such that $w_kw_{-k}<0$. 
\end{enumerate}
Then  there is $\delta >0$ such that for $|g|<\delta$ the spectrum of ${K}(g)$ contains at least a pair of complex conjugate  analytic arcs.
\end{theorem}
{\bf Remarks}
\begin{enumerate}
\item This theorem sharpens the results of \cite{Shin} in the sense that its  assumptions are explicit  because  they involve only the given  potential $W(x)$, while those of Theorem 3 and Corollary 4 of \cite{Shin} involve some conditions on  the Floquet discriminant of the equation   ${K}(g)\psi=E\psi$. This requires some a priori  information on the solutions of the equation itself.  \item
Explicit examples of potentials fulfilling the above conditions are: 
$$
W(x)=i\sin^{2k+1}{nx},  \quad k=0,1\ldots; \quad n\;{\rm odd}
$$
For $g=1$ these potentials have been considered in \cite{BDM}, where is is claimed that the spectrum is purely real.  A more careful examination by \cite{Shin} shows that the appearance of complex spectra cannot be excluded. 
\end{enumerate}
\section{Proof of the statements}
\setcounter{equation}{0}%
\setcounter{theorem}{0}%
Let us first state an elementary remark under the form of a lemma. Incidentally, this also proves Assertion 1 of Theorem 1.2. 
\begin{lemma}
\label{VPT}
Let $f(x)\in L^\infty(\R;\C)$ be $2\pi$ periodic, $f(x+2\pi)=f(x)$, $x\in\R$, and $PT$ symmetric, 
$\overline{f}(-x)=f(x)$. Consider its Fourier 
coefficients
$$
f_n=\frac{1}{2\pi}\int_{0}^{2\pi}f(x)e^{-inx}\,dx
$$
Then $\overline{f}_n=f_n$ $\forall\,n\in\Z$.
\end{lemma}
{\bf Proof}
\newline
The assertion is an immediate consequence  of the  Carleson-Fefferman theorem, which states the pointwise convergence of the Fourier expansion
$$
\overline{f}(-x)=\sum_{n\in\Z}\overline{f}_ne^{inx}=\sum_{n\in\Z}f_ne^{inx}=f(x)
$$
almost everywhere in $[0,2\pi]$. This proves the Lemma.
\par
To prove  Theorem 1.1, 
let us first recall that by the Floquet-Bloch theory (see e.g. \cite{BS},  \cite{Ea}),  $\lambda \in\sigma(H(g))$ if and only if the equation $H(g)\psi=\lambda \psi$ has a  non-constant bounded solution. In turn,  all bounded solutions  have the (Bloch) form
\be
\label{Bloch}
\psi_p(x;\lambda,g) =e^{ipx}\phi_p(x;\lambda,g)
\ee
where $p\in]-1/2,1/2]:=B$ (the Brillouin zone) and $\phi_p$ is $2\pi$-periodic.  It is indeed immediately checked that $\psi_p(x;\lambda,g)$ solves $H(g)\psi=\lambda\psi$ if and only if $\phi_p(x;\lambda,g)
$  is a solution of 
$$
H_p(g) \phi_p(x;\lambda,g)=\lambda \phi_p(x;\lambda,g).
$$
Here $H_p(g)$ is the operator in $L^2(0,2\pi)$  given by
\be
\label{Hpg}
H_p(g)u=\left(-i\frac{d}{dx}+p\right)^2u+qu+gWu, \quad u\in D(H_p(g))
\ee 
with periodic boundary conditions; its realization will be recalled below.  More precisely,  denote  $S^1$ the one-dimensional torus, i.e. the interval $[-\pi,\pi]$ with the endpoints identified. By Assumptions (1) and (2) above the restriction of $q$ to $S^1$, still denoted $q$ by a standard abuse of notation, belongs to $H^{-1}(S^1)$  and generates a real quadratic form $\Q_p(u)$ in $L^2(S^1)$ with domain  $H^{1}(S^1)$.  By  Assumption (3) $\Q_p(u)$ is  relatively bounded, with relative bound zero, with respect  to
\begin{eqnarray}
\T_p(u) := \int_{-\pi}^{\pi}[-iu^\prime +pu][i\overline{u}^\prime+p\overline{u}]\,dx, \quad D(\T_p(u))=H^1(S^1)
\end{eqnarray}
 so that the real semibounded  form $\H^0_p(u):=\T_p(u)+ \Q_p(u)$ defined  on $H^1(S^1)$ is closed. The corresponding self-adjoint operator in  $L^2(S^1)$ is the self-adjoint realization of the formal differential expression (note again the abuse of notation)
 $$
 H_p(0)=-\frac{d^2}{dx^2}-2ip\frac{d}{dx}+p^2+q
 $$
 As above, the form $\H_p(g)(u):=\H^0_p(u)+\la u,W u\ra$ defined on $H^1(S^1)$  is closed and sectorial in $L^2(S^1)$. Let ${ H}_p(g)$ the associated $m$-sectorial operator in $L^2(S^1)$. On $u\in D({ H}_p(g))$ the action of the operator  ${ H}_p(g)$ is specified by (\ref{Hpg}); moreover, $H_p(g)$ has compact resolvent.
\par\noindent
Let 
$$
\sigma(H_p(g)):=\{\lambda_n(g;p):\;n=0,1,\ldots\}
$$
 denote the spectrum of $H_p(g)$, with $p\in ]-1/2,1/2]$, $|g|<\overline{g}$. By the above remarks we have
 $$
 \sigma(H(g))=\bigcup_{p\in ]-1/2,1/2]}\sigma(H_p(g))=\bigcup_{p,n}\lambda_n(g;p)
 $$
  To prove the reality of $ \sigma(H(g))$, $|g|<\overline{g}$,  it is therefore enough to prove the reality of all eigenvalues $\lambda_n(g;p):\;n=0,1,\ldots; p\in ]-1/2,1/2]; n=0,1,\ldots$. 
 \par
  To this end, let us further recall the construction of the bands for $g=0$: it can be proved that  under the present conditions  all eigenvalues   $\lambda_{n}(0;p)$ are simple  $\forall\,p\in[-1/2,1/2]$;  the functions $\lambda_{n}(0;p)$  are continuous and even in  $[-1/2,1/2]$  with respect to $p$, so that  one can restrict to $p\in [0,1/2]$; the  functions $\lambda_{2k}(0;p)$ are strictly increasing on $[0,1/2]$ while the functions $\lambda_{2k+1}(0;p)$ are strictly decreasing, $k=0,1\ldots$.  Set:
$$
\alpha_k=\lambda_k(0,0),\quad \beta_k=\lambda_k(0,1/2)
$$
Then:
$$
\alpha_0<\beta_0<\beta_1<\alpha_1<\alpha_2< \beta_2<\beta_3 \ldots 
$$ 
The intervals $[\alpha_{2n},\beta_{2n}]$ and $[\beta_{2n+1},\alpha_{2n+1}]$ coincide with the range of $\lambda_{2n}(0,p)$, $\lambda_{2n+1}(0,p)$, respectively,  and represent the bands of  $\sigma(H(0))$; the intervals 
$$
\Delta_n:=]\beta_{2n},\beta_{2n+1}[,  \quad ]\alpha_{2n+1},\alpha_{2(n+2)}[
$$  
the gaps between the bands. 
\par\noindent
The monotonicity of the functions $\lambda_n(0,p)$ and Assumption (\ref{infgap}) entail
\be
\label{infband}
\inf_n \min_{p\in [0,1/2]} |\lambda_n(0,p)-\lambda_{n+1}(0,p)|\geq 2d
\ee
Let us now state the following preliminary result:
\begin{proposition}
\label{prop1}
\vskip 2pt\noindent
\begin{enumerate}
\item 
Let $g\in \overline{{\mathcal D}}$, where $\overline{\mathcal D}$ is the disk $\{g\,:\, |g|<\overline{g}\}$. For any $n$, there is a function $\lambda_n(g,p):\overline{\mathcal D}\times [0,1/2] \to \C$, holomorphic in $g$ and continuous in $p$, such that $\lambda_n(g,p)$ is a simple eigenvalue of ${\mathcal H}_p(g)$ for all $(g,p)\in \overline{\mathcal D}\times [0,1/2]$.
\item 
 $$
 \sup_{g\in \overline{\mathcal D}, p\in[0,1/2]}|\lambda_n(g,p)-\lambda_n(0,p)|<\frac{d}{2}
 $$
 \item If $g\in\R\cap \overline{\mathcal D}$ all eigenvalues $\lambda_n(g,p)$ are real;
 \item If $g\in\R\cap \overline{\mathcal D}$ then $\sigma({\mathcal H}_p(g))\equiv \{\lambda_n(g,p)\}_{n=0}^\infty$.
  \end{enumerate}
\end{proposition}
Assuming the validity of this Proposition  the proof of Theorem 1.1 is immediate.
\par\noindent
{\bf Proof of Theorem 1.1}
\newline
Since the functions $\lambda_n(g;p)$ are real and continuous for $g\in\R\cap \overline{\mathcal D}$, $p\in [0,1/2]$,  we can define:
\begin{eqnarray}
\label{estremi}
\alpha_{2n}(g):=\min_{p\in [0,1/2]}\lambda_n(g;p); \quad \beta_{2n}(g):=\max_{p\in [0,1/2]}\lambda_{2n}(g;p)\qquad
\\
\alpha_{2n+1}(g):=\max_{p\in [0,1/2]}\lambda_{2n+1}(g;p); \quad \beta_{2n+1}(g):=\min_{p\in [0,1/2]}\lambda_{2n+1}(g;p)\end{eqnarray}
Then:
$$
\sigma(H(g))=\bigcup_{n=0}^\infty B_n(g)
$$ 
where the bands $B_n(g)$ are defined, in analogy with the $g=0$ case, by:
\begin{eqnarray*}
B_{2n}(g)&:=&[\alpha_{2n}(g),\beta_{2n}(g)]
\\
B_{2n+1}(g)&:=&[\beta_{2n+1}(g),\alpha{2n+1}(g)]
\end{eqnarray*}
By Assertion 2 of Proposition 2.2 we have, $\forall\;n=0,1,\ldots$, $\forall\,g\in\overline{D}$:
\begin{eqnarray*}
\alpha_{2n}(g)-\frac{d}{2}=\lambda_{2n}(0,0)-\frac{d}{2}\leq \lambda_{2n}(0,p)-\frac{d}{2}
\\
\leq  \lambda_{2n}(g,p)\leq \lambda_{2n}(0,1/2)+\frac{d}{2}=\beta_{2n}+\frac{d}{2}\quad 
\end{eqnarray*}
whence
$$
\alpha_{2n}(g)-\frac{d}{2}\leq \lambda_{2n}(g,p)\leq \beta_{2n}+\frac{d}{2},\quad \forall\,n,\;\forall\,g\in\overline{\D}.
$$
This yields:
$$
B_{2n}(g)\subset \left[\alpha_{2n}-\frac{d}{2}
,  \beta_{2n}+\frac{d}{2}\right]
$$
By an analogous argument:
$$
B_{2n+1}(g)\subset \left[\beta_{2n+1}-\frac{d}{2}, \alpha_{2n+1}+\frac{d}{2}\right]
$$
Therefore the bands are pairwise disjoint, because the gaps 
$$
\Delta_n(g):= ]\beta_{2n}(g),\alpha_{2n}(g)[, \quad ]\alpha_{2n+1}(g),\beta_{2n+1}(g)[
$$
are all open and their width is no smaller than $d$. In fact, by (\ref{infgap}) we have: 
$$
|\alpha_n-\alpha_{n+1}|\geq 2d,\quad |\beta_n-\beta_{n+1}|\leq 2d .
$$
This concludes the proof of the Theorem
\vskip 0.3cm
We now prove separately the assertions of Proposition 2.2
\par\noindent
{\bf Proof of Proposition 2.2, Assertions 1 and 2}
\newline
Since the maximal multiplication  operator by $W$ is continuous in $L^2(S^1)$ with norm $\|W\|_\infty$,  the operator family $\H_p(g)$ is a type-A holomorphic family with respect to $g\in\C$, uniformly with respect to $p\in[0,1/2]$. Hence we can  direct apply regular perturbation theory (see e.g.\cite{Ka}): the perturbation expansion near any eigenvalue $\lambda_n(0;p)$ of $\H(0,p)$ exists and is convergent  in $\overline{\mathcal D}$ for $g\in \overline\D$ to a simple eigenvalue $\lambda_n(g;p)$ of $H(g;p)$:
\be
\label{ps}
\lambda_n(g;p)=\lambda_n(0,p)+\sum_{s=1}^\infty\lambda_n^s(0;p)g^s, \quad  g\in \overline\D
\ee
The convergence radius $r_n(p)$ is no smaller than $\overline{g}$. Hence $\overline{g}$ represents a lower bound for  $r_n(p)$ independent of $n$ and $p$. 
 Moreover $\lambda_n^s(0;p)$ is continuous for all $p\in[0,1/2]$, and hence  the same is true for the sum $\lambda_n(g;p)$.  This proves Assertion 1. 
\newline
To prove Assertion 2, recall that the coefficients  $\lambda_n^s(0;p)$ fulfill the majorization (see \cite{Ka}, \S II.3)
\be
\label{psm}
|\lambda_n^s(0;p)|\leq  \left(\frac{2\|W\|_\infty}{\inf_{k}\min_{[0,1/2]}|\lambda_k(0,p)-\lambda_{k\pm 1}(0,p)|}\right)^s\leq 
\left(\|W\|_{\infty}/d\right)^s
\ee
Therefore, by (\ref{ps}):
\begin{eqnarray*}
|\lambda_n(g;p)-\lambda_n(0,p)| &\leq& \frac{|g|\left(\|W\|_{\infty}/d\right)}{1-\left(2|g|\|W\|_{\infty}/d\right)}=\frac{|g|\|W\|_{\infty}}{d-|g|\|W\|_{\infty}} <  \frac{d}{2}
\end{eqnarray*}
 whence the stated majorization on account of (\ref{rc}).
\par\noindent
{\bf Proof of Proposition 2.2, Assertion 3}
\newline
As is known, and anyway very easy to verify,  the $PT$ symmetry entails that the eigenvalues of a  $PT$-symmetric operator are either real or complex conjugate. By standard regular perturbation theory (see e.g. \cite{Ka}, \S VII.2) any eigenvalue $\lambda_n(0;p)$ of $H_p(0)$ is stable with respect to $H_p(g)$;  since $\lambda_n(0,p)$ is simple, for $g$ suitably small there is one  and only one eigenvalue $\lambda_n(g,p)$ of $H_p(g)$  near $\lambda_n(0,p)$, and  $\lambda_n(g,p)\to \lambda_n(0,p)$ as $g\to 0$.  This excludes the existence of the complex conjugate eigenvalue $\overline{\lambda}_n(g,p)$ distinct from  $\lambda_n(g,p)$. Thus for $g\in\R$, $|g|$ suitably small, $\lambda_n(g,p)$ is real. This entails the reality of series expansion (\ref{ps}) for $g$ small and hence $\forall\,g\in\overline{\D}$. This in turn implies the reality of  $\lambda_n(g,p)\forall\,g\in \overline{\D}$.
\par\noindent
{\bf Proof of Proposition 2.2, Assertion 4}
\par\noindent
We repeat here the argument introduced in \cite{CGS2},\cite{CGC} to prove the analogous  result in different contexts. We describe all details to make the paper self contained.
We have seen that  for any $r\in \N$ the \RSPE\ associated with the  eigenvalue $\lambda_r(g;p)$ of $H_p(g)$ which converges to $\lambda_r(0;p)$ as $g\to 0$, has radius of convergence no smaller than $\overline{g}$. Hence, $\forall g\in \R$ such that $|g|<\overline{g}$, $H_p(g)$ admits a sequence of real eigenvalues $\lambda_r(g;p), r\in \N$.  We want to prove that for $|g|<\overline{g}, g\in\R$, $H_p(g)$ has no other eigenvalues.  Thus all its eigenvalues are real. To this end, for any $r\in\N$  let ${\cal Q}_r$ denote the circle centered at $\lambda_r(0;p)$ with radius $d$. 
 Then if $g\in\R$, $|g|<\overline{g}$, and $\lambda(g)$ is an eigenvalue of $H_p(g)$:
$$
\lambda(g)\in\bigcup_{r\in\N}\,{\cal Q}_r .
$$
In fact, denoting
$$
R_0(z):= (H_p(0) - z)^{-1}
$$
for any $\ds z\notin \bigcup_{r\in\N}\,{\cal Q}_r$ we have
\be
\label{Stima1}
\|gWR_0(z)\|\leq |g|\|W\|_\infty \|R_0(z)\|<\overline{g} \|W\|_\infty[{\rm dist}(z,\sigma(H_0))]^{-1}\leq \frac{\overline{g} \|W\|_\infty}{d}< 1 .
\ee
 The last inequality in (\ref{Stima1}) follows directly from the definition (\ref{rc}) of $\overline{g}$. Thus, $z\in\rho (H_p(g))$ and 
$$
R(g,z):= (H_p(g) - z)^{-1} = R_0(z)[1 + gWR_0(z)]^{-1}\,.
$$
Now let $g_0\in\R$ be fixed with $|g|<\overline{g} $. Without loss of generality we assume that $g_0>0$.  Let $\lambda(g_0)$ be a given eigenvalue of $H_p(g_0)$. Then  $\lambda(g_0)$ must be contained in the interior (and not on the boundary) of ${\cal Q}_{n_0}$ for some $n_0\in\N$. Moreover if $m_0$ is the multiplicity of $\lambda(g_0)$, for $g$ close to $g_0$ there are $m_0$ eigenvalues (counting multiplicities) $\lambda^{(s)}(g), s=1.\dots ,m_0$, of $H_p(g)$ which converge to $\lambda(g_0)$ as $g\to g_0$ and each function $\lambda^{(s)}(g)$ represents a branch of one or several holomorphic functions which have at most algebraic singularities at $g=g_0$ (see \cite{Ka}, Thm. VII.1.8). Let us now consider any one of such branches $\lambda^{(s)}(g)$ for $0<g<g_0$, suppressing the index $s$ from now on. First of all we notice that, by continuity, $\lambda(g)$ cannot lie outside ${\cal Q}_{n_0}$ for $g$ close to $g_0$. Moreover, if we denote $\Gamma_{t}$ the boundary of the circle centered at $\lambda_{n_0}(0;p)$ with radius $t$, $0<t\leq d$, we have, for $z\in\Gamma_{t}$ and $0<g\leq g_0$,
\be
\label{Stima2}
\|gWR_0(z)\|\leq g\|W\|_{\infty}[{\rm dist}(z,\sigma(H_p(0)))]^{-1}\leq g\|W\|_{\infty}/t \,.
\ee
Then $t>g\|W\|_{\infty}$ implies $z\notin \sigma(H_p(g))$, i.e. if $z\in \sigma(H_p(g))\cap\Gamma_{t}$ then $t\leq g\|W\|_{\infty}<g_0\|W\|_{\infty}<\overline{g}\|W\|_{\infty}<d$. Hence we observe that as $g\to g_0^-$, $\lambda(g)$ is contained in the circle centered at $\lambda_{n_0}(0;p)$ and radius $g\|W\|_{\infty}$. Suppose that the holomorphic function $\lambda(g)$ is defined on the interval $]g_1, g_0]$ with $g_1>0$. We will show that it can be continued up to $g=0$, and in fact up to $g=-\overline{g}$. From what has been established so far the function $\lambda(g)$ is bounded as $g\to g_1^+$. Thus, by the well known properties on the stability of the eigenvalues of the analytic families of operators, $\lambda(g)$ must converge to an eigenvalue $\lambda(g_1)$ of $H_p(g_1)$ as $g\to g_1^+$ and $\lambda(g_1)$ is contained in the circle centered at $\lambda_{n_0}(0;p)$ and radius $g_1\|W\|_{\infty}$. Repeating the argument starting now from $\lambda(g_1)$, we can continue $\lambda(g)$ to a holomorphic function on an interval $]g_2, g_1]$, which has at most an algebraic singularity at $g=g_2$. We build in this way a sequence $g_1>g_2>\dots >g_n>\dots $ which can accumulate only at $g=-\overline{g}$. In particular the function $\lambda(g)$ is piecewise holomorphic on  $]-\overline{g}, \overline{g}[$. But while passing through $g=0$, $\lambda(g)$ coincides with the eigenvalue $\lambda_r(g;p)$ generated by an unperturbed eigenvalue $\lambda_r(0;p)$ of $H_p(0)$ (namely $\lambda_{n_0}(0;p)$), which represents a real analytic function defined for $g\in ]-\overline{g},\overline{g}[$. Thus, $\lambda(g_0)$ arises from this function and is therefore real. This concludes the proof of Assertion $4$.
\vskip 0.3cm\noindent
{\bf Proof of Theorem 1.2}
\newline
Consider the operator $K_p(g)$ acting in $L^2(S^1)$, defined on the domain $H^2(S^1)$. By the Floquet-Bloch theory recalled above, we have again
$$
\sigma(K(g))=\bigcup_{p\in]-1/2,1/2]}\sigma(K_p(g)).
$$
It  is then enough to prove that there is $\eta>0$ such that $K_p(g)$ has complex eigenvalues for $p\in]1/2 -\eta,1/2]$.  Since $K_p(g)$ is $PT$-symmetric, eigenvalues may occur only in complex-conjugate pairs.  The eigenvalues  of $K_{p}(0)$ are $\lambda_n(0,p)=(n+p)^2: n\in\Z$.  The eigenvalue $\lambda_0(0,p)=p^2$ is simple $\forall\,p\in[0,1/2]$; any other eigenvalues is simple for $p\neq 0$, $p\neq 1/2$ and has multiplicity $2$ for $p=0$ or $p=1/2$ because $n^2=(-n)^2$ and $(n+1/2)^2=(-n-1+1/2)^2$, $n=0,1,\ldots$.   For $p=1/2$ a  set of orthonormal eigenfunctions corresponding to the double eigenvalue  $(n+1/2)^2=(-n-1+1/2)^2$, $n=0,1,\ldots$ is given by $\{u_n, u_{-n-1}\}$, where:
$$
u_n:=\frac1{\sqrt{2\pi}}e^{inx}, \quad n\in\Z
$$
Remark that,  for $0<p<1/2$,  $u_n$ and $u_{-n-1}$ are the (normalized) eigenfunctions corresponding to the simple eigenvalues $(n+p)^2$, $(-n-1+p)^2$.   Let now $n=k\in\N$, and consider the $2\times 2$ matrix 
$$
T:=\left(\begin{array}{ll} \la u_k,Wu_k\ra & \la u_k,Wu_{-k-1}\ra 
\\ \la u_{-k-1},Wu_k\ra  & \la u_{-k-1},Wu_{-k-1}\ra \end{array}\right)
$$ 
A trivial computation yields
$$
T:=\left(\begin{array}{ll} 0 &  w_{2k+1}
\\ w_{-2k-1}  & 0 \end{array}\right)
$$
with purely imaginary eigenvalues $\ds \mu^\pm=\pm i\sqrt{-w_{2k+1}w_{-2k-1}}$. 
By standard degenerate perturbation theory, for $g$ small enough $\H_{1/2}(g)$ admits the  pair of complex conjugate eigenvalues
\be
\label{cev}
\lambda_k^\pm(g,1/2)=(k+1/2)^2 \pm i g \sqrt{-w_{2k+1}w_{-2k-1}}+O(g^2)
\ee
Under the above assumptions, for any fixed $g\in\R$ the operator family $p\mapsto K_p(g)$ is type-A holomorphic in the sense of Kato (see \cite{Ka}, \S VII.2) $\forall\,p\in\C$  because its domain  does not depend on $p$ and the scalar products $\la u,K_p(g)u\ra:  u\in H^1(S^1)$ are obviously   holomorphic functions    of $p$ $\forall\,p\in \C$.  This entails the continuity with respect to $p$  of the   eigenvalues 
$\lambda_n(g;p)$.  Therefore, for $|g|$ suitably small, there is $\eta(g)>0$ such that 
$$
{\rm Im}\,\lambda_k^\pm(g,p)\neq 0, \quad 1/2-\eta \leq p\leq 1/2
$$
It follows (see e.g.\cite{BS}, \cite{Ea}) that the complex  arcs ${\mathcal E}^\pm_k:={\rm Range}(\lambda_k^\pm(g,p)): p\in[1/2-\eta(g), 1/2]$ belong to the spectrum of ${ K}(g)$. This concludes the proof of the theorem.

\end{document}